\documentclass[aps,prl,preprint]{revtex4-1}  
\usepackage{graphicx}  
\usepackage{dcolumn}   
\usepackage{bm}        
\usepackage{amssymb}   
\usepackage{xcolor}

\newcommand{\be}{\begin{equation}}
\newcommand{\ee}{\end{equation}}
\newcommand{\ba}{\begin{eqnarray}}
\newcommand{\ea}{\end{eqnarray}}

\definecolor{verde}{cmyk}{.83,.21,1,.08}

\usepackage{ulem}

\begin{document}

\preprint{ICCUB-16-012, KCL-PH-TH/2016-12}

\author{Marco de Cesare$^1$, Fedele Lizzi$^{2,3,4}$ and Mairi Sakellariadou$^1$}

\affiliation{$^1$Department of Physics, King's College London, University of London, 
London, United Kingdom}
\affiliation{
$^2$Dipartimento di Fisica ``E.\ Pancini'', Universit\`{a} di Napoli ``Federico II'', 
Napoli, Italy}
\affiliation{$^3$INFN, Sezione di Napoli, Italy}
\affiliation{$^4$Departament de Estructura i Constituents de la Mat\`eria, Institut de Ci\'encies del Cosmos, Universitat de Barcelona}

\email{marco.de\_cesare@kcl.ac.uk}
\email{fedele.lizzi@na.infn.it}
\email{mairi.sakellariadou@kcl.ac.uk}

\title{Effective 
cosmological constant
induced by stochastic fluctuations of Newton's constant}

\begin{abstract}
We consider implications of the microscopic dynamics of spacetime for
the evolution of cosmological models. We argue that quantum geometry
effects may lead to stochastic fluctuations of the gravitational
constant, which is thus considered as a macroscopic effective dynamical
quantity. Consistency with Riemannian geometry entails the presence of a
time-dependent dark energy term in the modified field equations, which
can be expressed in terms of the dynamical gravitational constant.  We
 suggest that the late-time accelerated expansion of the Universe
may be ascribed to quantum fluctuations in the geometry of spacetime
rather than the vacuum energy from the matter sector.
\end{abstract}

\maketitle 

It is unlikely that the geometry of spacetime remains
unchanged all the way to microscopical scales, smaller than the
Panck's length $l_{\rm Pl}=\sqrt{\hbar G_{\rm N}/c^3}$, where quantum
effects can no longer be neglected and geometry may altogether lose
the meaning we are familiar with. The spacetime structure at
microscopic scales is of course not known, although some indications
from Noncommutative Geometry~\cite{Connes,WSS}, String
Theory~\cite{String}, or Loop Quantum Gravity~\cite{LQG} are
available.  Even though a full theory of quantum spacetime is not yet
available, it is reasonable to assume that Einstein's equations (as
well as the ones of quantum field theory) will be substituted by a new
set of relations. We may in
the meantime try to surmise some sort of effects which could be the
outcome of the quantum substructure. The aim of this paper is to
investigate the possibility that the coupling between matter and the
gravitational field varies in a random way, possibly due to quantum
gravitational effects.

The variability of fundamental constants is an old idea which goes
back to Dirac~\cite{Dirac}, and it would be a clear indication of new
physics. In fact, the variation of fundamental constants would entail
a violation of the Equivalence Principle~\cite{Uzan, Will} and
could not find an explanation within the same framework where those
constants belong. In the language of effective field theories, such
constants would be replaced by new (scalar) fields obeying their own
dynamics~\cite{Uzan}, as for instance in the framework of Brans-Dicke
theory~\cite{BransDicke} (or more general scalar-tensor theories of
gravity), where the gravitational constant is determined by the vacuum
expectation value of the dilaton. Clearly, in this framework the variation of
fundamental constants must be slow enough, both in space and time, so
as to guarantee the reproducibility of the experiments performed so
far. The possibility of measuring such variations, as well as the
theoretical reasons for introducing such dimensionful quantities in
first place, are actually delicate issues that have generated debate
in the community. Interested readers are referred to e.g.\
Refs.~\cite{Duff, Trialogue}. We will see in the following
that the variability of the
coupling, which can be expressed in a unit independent fashion in
terms of dimensionless quantities, is the physically relevant (and measurable) quantity. While most of the theoretical work
on variable constants has been on a more or less regular variability
on cosmological scales, the motivations of this work point to a
\textit{stochastic} variability on \textit{very short scales}, typically
Planckian.  Thus, it is in general conceivable that $G_{\rm N}$ could
not be a constant at the outset but instead it must be determined from
the microscopic dynamics, hence in principle it can be affected by the
collective behaviour of spacetime quanta. Analogous considerations
apply to $\hbar$~\cite{ManganoLizziPorzio} for the effects
induced by its stochastic fluctuations on nonrelativistic quantum
systems. In fact, it is a general property of quantum systems that
fast degrees of freedom can be approximated by stochastic noise, see
Ref.~\cite{book noise}. In the following, we assume that the effects
of the underlying microscopic dynamics of spacetime can be given an
\textit{effective classical description} by means of a stochastic
gravitational constant. The fluctuations and the typical time scale of
the stochastic process must be suitably small, \textit{e.g.}
${\cal {O}}(t_{\rm Pl})$, to ensure that there is no contradiction with
observations. A fluctuating gravitational background can
induce the Higgs mass term as well~\cite{Kurkov}.

Newton's gravitational constant appears in the fundamental equation of Einstein's theory of General Relativity
\be\label{eq:Einstein}
G_{ab}=8\pi G_{\rm N}\; T_{ab}~.
\ee
We take Eq.~(\ref{eq:Einstein}) as a starting point and work out the
consequences of having a dynamical $G_{\rm N}$. Turning $G_{\rm N}$ naively into a
dynamical variable, while keeping $T_{ab}$ covariantly conserved, would
result into a violation of the Bianchi identities, as pointed out
already in Ref.~\cite{Fritzsch}. 
 \be\label{eq:Bianchi}
0=\nabla^{a}G_{ab}=8\pi \nabla^{a}(G_{\rm N} T_{ab})~.  \ee 
The solution to
the impasse is to split the energy-momentum tensor into a matter
term, which satisfies the usual conservation law, and a correction
term such that Eq.~(\ref{eq:Bianchi}) is satisfied:
\be
T_{ab}=T^{\rm matter}_{ab}+\tau_{ab}~.  \ee
 In a isotropic and
homogeneous cosmological model, we have \be
T^{\rm matter}_{ab}=(\rho+p)u_{a}u_{b}+p g_{ab}~, \ee where $u_a$ is
tangent to the geodesic of a isotropic observer and the fluid
satisfies the equation of state $p=w\rho$.  A similar ansatz could be
made for the extra term. However, since it represents the only part of
the energy-momentum tensor that does not arise from the matter sector,
it is quite natural to connect it to the only such form of energy
there is evidence for, namely to {\sl  dark energy}. We therefore write
\be\label{eq:cosmological constant} \tau_{ab}=-\lambda g_{ab}~.  \ee In
this way dark energy is naturally connected to the dynamics of $G_{\rm N}$,
stemming from quantum gravitational effects. There is no need for other
physical fields to be introduced in the theory. In contrast to
Ref.~\cite{Fritzsch}, we are not trying to establish a connection with
the matter sector in order to explain such variations. Rather, we
argue that their origin is entirely due to quantum geometry
effects. At the macroscopic level, these must be such that general
covariance and the Bianchi identities are preserved. An extra
constraint on the form of $\tau_{ab}$ comes from observations, which
tell us that it has to satisfy Eq.~(\ref{eq:cosmological constant}).
In the following we will consider just variation over time. This is to be considered a toy model of a full treatment for which variations are, always at Planckian scales, in both space and time.

The physical fluid satisfies the continuity equation
\be\label{eq:continuity}
\dot{\rho}+3H(\rho+p)=0~.
\ee
The evolution of the time-dependent ``cosmological constant''
$\lambda$ is obtained from the Bianchi identities
\be
\dot{G}_{\rm N}(\rho+\lambda)+G_{\rm N}\dot{\lambda}=0~.
\ee
The solution to this equation is
\be\label{eq:solution lambda} \lambda(t)=\frac{G_{\rm N}(t_i)}{G_{\rm
    N}(t)}\left(\lambda(t_i)-\int_{t_i}^{t}\mbox{d}t^{\prime}\;\frac{\dot{G}_{\rm
    N}(t^{\prime})}{G_{\rm N}(t_i)}\rho(t^{\prime})\right)~.  \ee
By substituting Eq.~(\ref{eq:solution lambda}) in the sources of the
Friedmann equation we get
\be\label{eq:Friedmann}
H^2(t)=\frac{8\pi}{3}\left(G_N(t_i)\left(\rho(t_i)+\lambda(t_i)\right)+\int_{t_i}^{t}G_N\dot{\rho}\right)~.
\ee
Considering $t_i$ in the matter- or radiation-dominated era,
$\lambda(t_i)\ll\rho(t_i)$ and we can safely neglect the second term
in the r.h.s.\ of the above equation. Notice that in this case,
Eq.~(\ref{eq:Friedmann}) does not make any reference to $\lambda$, but all
dark energy contributions are instead included in the time
evolution of $G_{\rm N}$ via an integral operator.

By differentiating Eq.~(\ref{eq:Friedmann}) and using
Eq.~(\ref{eq:continuity}), in the case of a Universe filled with
non-relativistic matter ($w=0$), we get
\be\label{eq:acceleration}
\dot{H}=-4\pi G_{\rm N} \rho~, \ee 
which together with Eq.~(\ref{eq:continuity})
and the constraint on the initial conditions for $\rho$ and $H$, namely
\be\label{eq:initial constraint}
H^2(t_i)=\frac{8\pi}{3}G_{\rm N}(t_i)\rho(t_i)~, \ee 
fully determines the
dynamics. We can express 
\be\label{eq:stochastic G}
G_{\rm N}(t)=\overline{G}_{\rm N}(1+\sigma\xi(t)) \ee
 as the sum of a constant background value and a perturbation. Note
 that the last equation still makes sense even if one assumes the
 point of view of Ref.~\cite{Duff}. In fact, once we introduce a units
 system, \textit{e.g.}  one such that $\overline{G}_{\rm N}=1$,
 Eq.~(\ref{eq:stochastic G}) is equivalent to the statement that the
 r.h.s.\ of Einstein's equations (\ref{eq:Einstein}) has a nonstandard
 form, with new terms that can be expressed in terms of dimensionless
 quantities.

Let us rewrite Eq.~(\ref{eq:acceleration}) in a differential form, namely
\be
\mbox{d}H=-4\pi \overline{G}_{\rm N} \rho (1+\sigma\xi(t))\mbox{d}t~.
\ee
Assuming that $\xi(t)$ has a white noise distribution, the last
equation reads 
\be\label{eq:Ito} \mbox{d}H=-4\pi
\overline{G}_{\rm N} \rho (\mbox{d}t+\sigma\mbox{d}W_t)~, \ee 
where $W_t$ is a Wiener process. Loosely speaking the white noise is
the derivative of the Wiener process and satisfies the properties
\begin{eqnarray}
\langle\xi(t)\rangle&=&0,\\
\langle\xi(t)\xi(t^{\prime})\rangle&=&\delta(t-t^{\prime})~,
\end{eqnarray}
where $\langle~\rangle$ denotes an ensemble average.  A mathematically
more precise description of a stochastic process is given by
considering a \textit{discrete time evolution} and using the theory of
stochastic integration~\cite{Oksendal}. Thus,
Eq.~(\ref{eq:Ito}) is interepreted as a {\^I}to process.
For the sake of simplicity we consider a partition of the interval
$t_i\equiv t_0 <t_1<\dots <t_k<\dots<t_N\equiv t$
with uniform time step $h$.

The Wiener process $W_t$ has the following properties:
\begin{itemize}
\item $W_{t_i}=0$ with probability $1$;
\item The increments $W_{t_{k+1}}-W_{t_{k}}$ are statistically independent Gaussian variables
with mean $0$ and variance $t_{k+1}-t_{k}=h$.
\end{itemize}
It plays the r{\^o}le of a stochastic driving term in Eq.~(\ref{eq:Ito}).

\begin{figure}
\includegraphics[width=\columnwidth]{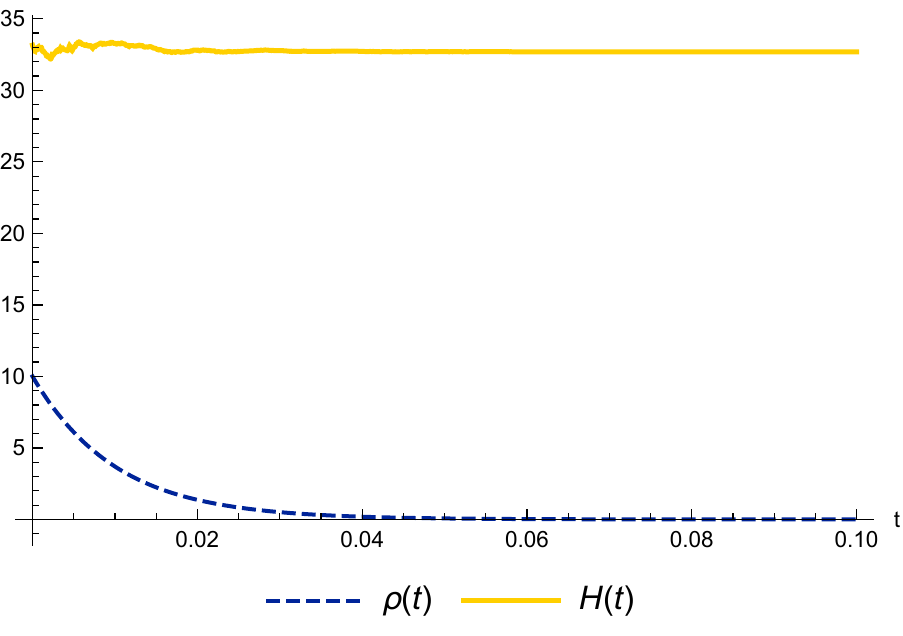}
\caption{\sl Plot of the solution of the model (\ref{eq:continuity}), (\ref{eq:initial constraint}), (\ref{eq:Ito}), for the particular choice of parameters $\sigma=0.1$, $\rho_0=10$ and time step $h=10^{-4}$, in units
  such that $\overline{G}_{\rm N}=1$. The horizontal asymptote of $H(t)$ for $t\to\infty$ corresponds to a late era of accelerated expansion, when the Universe is dominated by a ``cosmological constant''. Stochastic noise has an effect which is relevant at early times, whereas it is negligible at late times since it is suppressed by the small value of the energy density, as one can also see from (\ref{eq:Ito}). The qualitative behaviour of the solutions is a general feature, which does not depend on the particular choice of parameters.}\label{fig:h rho}
\end{figure}

We have investigated numerically the behaviour of the solutions for such a model with parameters $0\leq\sigma\leq1$, $10^{-6}\leq h\leq 10^{-2}$, in units such that $\overline{G}_{N}=1$, for different choices of the initial condition $\rho(t_i)$.
Qualitatively the behaviour of the solutions  is the same and is represented in Fig.~\ref{fig:h rho} for a given path $W_t$, where the plot of the energy density $\rho(t)$
and the Hubble rate $H(t)$ is shown.

As $t\to\infty$, $\rho$ decays as in standard cosmology, while $H$
attains a nonvanishing positive limit. The value of the asymptotic limit does not vary much with the random sequence, and is not particularly sensitive to the parameters. Since we are dealing with a simplified model this semi qualitative analysis is sufficient. This implies an exponential
expansion at late times without the need to introduce a cosmological
constant by hand or to originate it from the matter sector,
\textit{e.g.} as vacuum energy. Note that this is a \textit{general feature
of the model} that does not depend
on the particular values chosen for $\rho_0$ or the noise strength
$\sigma$. Furthermore, when $\sigma=0$ one recovers the standard
cosmology in the matter-dominated era, namely $\rho
\propto 1/t^2$ and $H \propto 1/t$.

The effective cosmological constant in the r.h.s.\ of
Eq.~(\ref{eq:Friedmann}) is given by the formula 
\be\label{eq:effective Lambda}
\Lambda_{\rm eff}(t)=\sigma\left(\xi(t_i)\rho(t_i)+\int_{t_i}^{t}\dot{\rho}\;
\mbox{d}W_t\right)~, \ee 
where the second term is a stochastic
integral. The first term dominates over the second with probability close to $1$,
as shown in the following. Hence, when the first step of the
random walk is positive, the Universe is exponentially expanding at
late times, whereas for a negative initial step $H$ takes large
negative values. We are therefore led to assume the initial condition
\be\label{eq:initial noise}
\xi(t_i)>0~, \ee 
which guarantees the stability of the Universe and is compatible with
the observed accelerated expansion.

We compute the probability to observe a negative value of the
effective cosmological constant given the initial condition on the noise (\ref{eq:initial noise}).
This actually forces the first step of the random walk to follow a half-normal
distribution, whereas all the other steps of the random walk
are normally distributed, as it is customary.
Using the statistical properties of the increments of the Wiener process
$W_t$ one finds
\ba
\left\langle \frac{\Lambda(t)}{\sigma}\right\rangle&=&\left\langle \frac{W_{t_i}}{h}\right\rangle\rho(t_i)=\sqrt{\frac{2}{\pi h}}\rho(t_i)\\
\left\langle \frac{\Lambda^2(t)}{\sigma^2}\right\rangle&=&\left\langle \left(\frac{W_{t_i}}{h}\right)^2\right\rangle\rho^2(t_i)+\sum_k (\rho^{\prime}(t_k))^2 h+2\rho(t_i)\rho^{\prime}(t_i)\nonumber\\
&\approx&\left\langle \left(\frac{W_{t_i}}{h}\right)^2\right\rangle\rho^2(t_i)=\frac{\rho^2(t_i)}{h}
\ea
Therefore the value $\Lambda=0$ is $x$ standard deviations away from its mean value,
with
\be
x=\sqrt\frac{2\pi}{\pi-2}.
\ee
Hence, the probability of observing a non-positive value of $\Lambda$
(leading to a collapsing Universe) is $\lesssim0.0095$, regardless
of the initial condition, the noise strength $\sigma$ and the time step.
In a more complete treatment which may take also space variation into account, these unstable cases will anyway not be a problem, since the instabilities will be just covered by the expanding areas.

Considering a Universe with more than one matter component, the
effective cosmological constant receives contribution from each of
them, with the dominant contribution still coming from the initial
value of the white noise $\xi(t_i)$ multiplied by the \textit{total energy density}
as in (\ref{eq:effective Lambda}). In other words, the effective cosmological
constant depends on the initial value of the total energy density,
but not on the species populating the Universe.
It seems natural to fix the initial data at a time where all species
where equally dominating, \textit{i.e.} $t_i\approx t_{\rm Pl}$.
 This conclusion, if correct, due to the
limits of our effective approach at Planckian times, implies that the
final stage of evolution of the Universe is entirely determined by
quantum fluctuations of the spacetime geometry at early
times. Furthermore, being determined only by the initial
value of the total energy density, it treats all fields on the same level
and it is insensitive to further details of the Universe's history.
 In this sense Eq.~(\ref{eq:initial noise}) can be interpreted
 as a constraint on the underlying quantum
theory, at the time when the dynamics of the fast degrees of freedom
of the gravitational field approach their stochastic limit.

\section{Outlook and conclusions}
Within an effective macroscopic description of the dynamics of the
gravitational field, we have considered a stochastically fluctuating
gravitational constant as a possible phenomenological feature of a
theory of quantum gravity. Consistency with the Bianchi identities
requires the addition of an extra source term, which does not
communicate with the matter sector and may be interpreted as  dark
energy. We built a simple model where the latter is seen to be
responsible for the present acceleration in the
expansion of the Universe, which is thus seen to have a purely quantum
geometrical origin. 
The simplifications introduced in the model and the sensitivity
of the effective cosmological constant to the value of the total energy density
at the Planck time
prevent us from using such model to
extract quantitative predictions for the present value of the cosmological
constant and its probability.
It is our hope that this preliminary work could open
new ways towards a solution to the cosmological constant problem.

If the picture obtained from our model could be consistently obtained
from a fundamental theory of quantum gravity, it would show that
quantum gravitational phenomena are already being observed in the
Universe today when looking at the acceleration of its expansion rate.
Moreover, it would
provide indirect evidence for the variation over time of the
gravitational constant. More work has to be done in order to gain a better
understanding of the r{\^o}le played by the initial conditions
on the energy content of the Universe,
since the cosmological constant obtained from our toy
model seems to have a strong dependence on them.

\noindent\textbf{Acknowledgments}
This article is based upon work from COST Action MP1405 QSPACE, supported by COST (European Cooperation in Science and Technology).  M.d.C. would like to thank Tobias Hartung for helpful discussions. M.d.C. and F.L. also thank G.Mangano for useful discussions.
F.L.\ is supported by INFN, I.S.'s GEOSYMQFT and received partial support by CUR Generalitat de Catalunya under projects FPA2013-46570 and 2014~SGR~104, MDM-2014-0369 of ICCUB (Unidad de Excelencia `Maria de Maeztu').

\end{document}